\documentclass[]{eptcs}
\providecommand{\event}{TFPIE 2019} %
\usepackage{underscore}           %

\usepackage[utf8]{inputenc}
\usepackage{amsmath}
\usepackage{amsfonts}
\usepackage{amssymb}
\usepackage{graphicx}

\usepackage{mathtools}

\usepackage{stackengine}
\stackMath

\usepackage{bussproofs}

\usepackage{relsize}

\usepackage{xcolor}

\usepackage{float}
\floatstyle{boxed}
\restylefloat{figure}

\usepackage{enumitem}

\usepackage{textpos}

\makeatletter
\@ifundefined{lhs2tex.lhs2tex.sty.read}%
  {\@namedef{lhs2tex.lhs2tex.sty.read}{}%
   \newcommand\SkipToFmtEnd{}%
   \newcommand\EndFmtInput{}%
   \long\def\SkipToFmtEnd#1\EndFmtInput{}%
  }\SkipToFmtEnd

\newcommand\ReadOnlyOnce[1]{\@ifundefined{#1}{\@namedef{#1}{}}\SkipToFmtEnd}
\usepackage{amstext}
\usepackage{amssymb}
\usepackage{stmaryrd}
\DeclareFontFamily{OT1}{cmtex}{}
\DeclareFontShape{OT1}{cmtex}{m}{n}
  {<5><6><7><8>cmtex8
   <9>cmtex9
   <10><10.95><12><14.4><17.28><20.74><24.88>cmtex10}{}
\DeclareFontShape{OT1}{cmtex}{m}{it}
  {<-> ssub * cmtt/m/it}{}

\DeclareFontShape{OT1}{cmtt}{bx}{n}
  {<5><6><7><8>cmtt8
   <9>cmbtt9
   <10><10.95><12><14.4><17.28><20.74><24.88>cmbtt10}{}
\DeclareFontShape{OT1}{cmtex}{bx}{n}
  {<-> ssub * cmtt/bx/n}{}

\newcommand{\Conid}[1]{\mathit{#1}}
\newcommand{\Varid}[1]{\mathit{#1}}
\newcommand{\anonymous}{\kern0.06em \vbox{\hrule\@width.5em}}
\newcommand{\plus}{\mathbin{+\!\!\!+}}
\newcommand{\bind}{\mathbin{>\!\!\!>\mkern-6.7mu=}}
\newcommand{\sequ}{\mathbin{>\!\!\!>}}
\renewcommand{\leq}{\leqslant}

\usepackage{polytable}

\@ifundefined{mathindent}%
  {\newdimen\mathindent\mathindent\leftmargini}%
  {}%

\def\resethooks{%
  \global\let\SaveRestoreHook\empty
  \global\let\ColumnHook\empty}
\newcommand*{\savecolumns}[1][default]%
  {\g@addto@macro\SaveRestoreHook{\savecolumns[#1]}}
\newcommand*{\restorecolumns}[1][default]%
  {\g@addto@macro\SaveRestoreHook{\restorecolumns[#1]}}
\newcommand*{\aligncolumn}[2]%
  {\g@addto@macro\ColumnHook{\column{#1}{#2}}}

\resethooks

\newcommand{\onelinecommentchars}{\quad-{}- }
\newcommand{\commentbeginchars}{\enskip\{-}
\newcommand{\commentendchars}{-\}\enskip}

\newcommand{\visiblecomments}{%
  \let\onelinecomment=\onelinecommentchars
  \let\commentbegin=\commentbeginchars
  \let\commentend=\commentendchars}

\newcommand{\invisiblecomments}{%
  \let\onelinecomment=\empty
  \let\commentbegin=\empty
  \let\commentend=\empty}

\visiblecomments

\newlength{\blanklineskip}
\newcommand{\hsindent}[1]{\quad}%
\let\hspre\empty
\let\hspost\empty

\EndFmtInput
\makeatother
\ReadOnlyOnce{polycode.fmt}%
\makeatletter

\newcommand{\hsnewpar}[1]%
  {{\parskip=0pt\parindent=0pt\par\vskip #1\noindent}}

\newcommand{\hscodestyle}{}

\newcommand{\sethscode}[1]%
  {\expandafter\let\expandafter\hscode\csname #1\endcsname
   \expandafter\let\expandafter\endhscode\csname end#1\endcsname}

  {\par\noindent
   \advance\leftskip\mathindent
   \hscodestyle
   \let\\=\@normalcr
   \let\hspre\(\let\hspost\)%
   \pboxed}%
  {\endpboxed\)%
   \par\noindent
   \ignorespacesafterend}

  {\hsnewpar\abovedisplayskip
   \advance\leftskip\mathindent
   \hscodestyle
   \let\hspre\(\let\hspost\)%
   \pboxed}%
  {\endpboxed%
   \hsnewpar\belowdisplayskip
   \ignorespacesafterend}

  {\hsnewpar\abovedisplayskip
   \advance\leftskip\mathindent
   \hscodestyle
   \let\\=\@normalcr
   \(\pboxed}%
  {\endpboxed\)%
   \hsnewpar\belowdisplayskip
   \ignorespacesafterend}

\newcommand{\plainhs}{\sethscode{plainhscode}}

\plainhs

  {\hsnewpar\abovedisplayskip
   \advance\leftskip\mathindent
   \hscodestyle
   \let\\=\@normalcr
   \(\parray}%
  {\endparray\)%
   \hsnewpar\belowdisplayskip
   \ignorespacesafterend}

  {\parray}{\endparray}

  {\(\parray}{\endparray\)}

\def\codeframewidth{\arrayrulewidth}
\RequirePackage{calc}

  {\parskip=\abovedisplayskip\par\noindent
   \hscodestyle
   \arrayrulewidth=\codeframewidth
   \tabular{@{}|p{\linewidth-2\arraycolsep-2\arrayrulewidth-2pt}|@{}}%
   \hline\framedhslinecorrect\\{-1.5ex}%
   \let\endoflinesave=\\
   \let\\=\@normalcr
   \(\pboxed}%
  {\endpboxed\)%
   \framedhslinecorrect\endoflinesave{.5ex}\hline
   \endtabular
   \parskip=\belowdisplayskip\par\noindent
   \ignorespacesafterend}

\newcommand{\framedhslinecorrect}[2]%
  {#1[#2]}

  {\(\def\column##1##2{}%
   \let\>\undefined\let\<\undefined\let\\\undefined
   \newcommand\>[1][]{}\newcommand\<[1][]{}\newcommand\\[1][]{}%
   \def\fromto##1##2##3{##3}%
   }{\) }%

  {\let\orighscode=\hscode
   \let\origendhscode=\endhscode
   \def\endhscode{\def\hscode{\endgroup\def\@currenvir{hscode}\\}\begingroup}
   \orighscode\def\hscode{\endgroup\def\@currenvir{hscode}}}%
  {\origendhscode
   \global\let\hscode=\orighscode
   \global\let\endhscode=\origendhscode}%

\makeatother
\EndFmtInput

\newcommand{\readInput}[2]{\ensuremath{[\, \triangleright\, #1 \,]^{#2}}}
\newcommand{\writeOutputSimple}[1]{\ensuremath{[\,#1\, \triangleright\, ]}}
\newcommand{\writeOutput}[1]{\ensuremath{[\,\{#1\}\, \triangleright\, ]}}
\newcommand{\restCh}[1]{\ensuremath{\mathop{\chAngle #1 \revChAngle}}}
\newcommand{\chAngle}{\raisebox{-1pt}{$\boldsymbol{\angle}$}}
\DeclareRobustCommand{\revChAngle}{\text{\reflectbox{\chAngle}}}
\newcommand{\nat}{\ensuremath{\mathbb{N}}}
\newcommand{\integer}{\ensuremath{\mathbb{Z}}}
\newcommand{\loopArr}{\ensuremath{{\rightarrow^\mathbf{E}}}}
\newcommand{\nop}{\ensuremath{{\mathbf{0}}}}
\newcommand{\loopExit}{\ensuremath{{\mathbf{E}}}}

\newcommand{\traceStop}{\mathit{stop}}

\newcommand{\accept}{\operatorname{\mathit{accept}}}

\newcommand{\traceIn}[1]{\mathop{?{#1}}}
\newcommand{\traceOut}[1]{\mathop{!{#1}}}
\newcommand{\traceVar}[1]{\,#1}

\newenvironment{adjustedcenter}{\begin{center}\vspace{-4ex}}{\vspace{-1ex}\end{center}}

\title{Describing Console I/O Behavior for Testing Student Submissions in Haskell}
\author{Oliver Westphal \qquad\qquad Janis Voigtländer
\institute{Universität Duisburg-Essen\\
Germany}
\email{\quad oliver.westphal@uni-due.de \quad\qquad janis.voigtlaender@uni-due.de}
}

\def\titlerunning{Describing I/O Behavior}
\def\authorrunning{O. Westphal, J. Voigtländer}

\begin{document}
\maketitle

\begin{abstract}
We present a small, formal language for specifying the behavior of simple console I/O programs.
The design is driven by the concrete application case of testing interactive Haskell programs written by students.
Specifications are structurally similar to lexical analysis regular expressions, but are augmented with features like global variables that track state and history of program runs, enabling expression of an interesting range of dynamic behavior.
We give a semantics for our specification language based on acceptance of execution traces.
From this semantics we derive a definition of the set of all traces valid for a given specification.
Sampling that set enables us to mechanically check program behavior against specifications in a probabilistic fashion.
Beyond testing, other possible uses of the specification language in an education context include related activities like providing more helpful feedback, generating sample solutions, and even generating random exercise tasks.
\end{abstract}

\section{Introduction}

In our course on programming paradigms we teach the main concepts of Haskell.
For students to gain practical experience with the language, we give weekly exercise tasks and let them submit solutions for review and grading.
Since checking submissions by hand is tedious and potentially error-prone, we employ an e-learning system~\cite{DBLP:conf/abp/SiegburgVW19,DBLP:conf/abp/Waldmann17}, automatically testing submissions against sets of QuickCheck properties~\cite{claessen2000quickcheck} wherever possible.
An added benefit for students is that they get immediate feedback and can revise their submissions accordingly and incrementally.

This approach works very well for exercise tasks about implementing pure functions.
But applying it to tasks about I/O programming is significantly more complicated.
While Swierstra and Altenkirch~\cite{swierstra2007} showed how one can change the monad underlying a Haskell I/O program in order to get an inspectable representation of what the program is doing, thus in principle enabling the checking of properties formulated over executions via QuickCheck or similar tools, practical application is cumbersome.
Specifically, for every I/O exercise task we want to handle that way, we currently need to implement the following components:
\begin{itemize}[noitemsep]
  \item A generator of input sequences, respecting the task's invariants.
  \item A way of checking if a given execution trace exhibits the desired behavior for a given input sequence.
  \item A method of providing feedback in case the behavior did not match the expectations, e.g., explaining the mismatch or showing what would be correct behavior for the relevant input sequence.
\end{itemize}
Since we typically allow students some degree of freedom when it comes to how their program should prompt for input values, in what form exactly it should print any computed result values, whether there are additional/optional output messages, etc., these components have to cover a lot of different cases even for a single exercise task.
Moreover, the components are not directly related to each other, much less derived from a common source, thus leaving room for inconsistencies and other errors.
We lack an overall framework, and it shows.

Our main idea here is that much of this complexity can be tamed by designing a domain-specific language for specifying interactive program behavior in a way that enables automatic generation of the components listed above.
We therefore make the following contributions:
\begin{itemize}[noitemsep]
  \item We design (Section~\ref{sec:design}), then describe the syntax (Section~\ref{sec:syntax}) and semantics (Section~\ref{sec:semantics}), of a small language suitable for specifying simple programming tasks about console I/O.
  Key features are history aware variables that allow access to all values previously read into them and an encoding of optional or ambiguous output behavior.
  \item
  For any given specification, we show how to generate generalized traces that cover the complete behavior accepted by that specification under a fixed sequence of inputs.
  Based on this trace generation process we can build the three necessary components, mentioned above, for automatically testing exercise tasks (Section~\ref{sec:testing}).
  \item
  We start to implement the presented approach via an embedded domain-specific language (EDSL) in Haskell (Section~\ref{sec:impl}).
  More details and further development are reported in a companion paper~\cite{flops20-paper}.
\end{itemize}
For the sake of simplicity of the described formalism, we will only consider programs that read and write integers.
It is however straightforward to generalize the approach to, for example, include string values, and the implemented EDSL actually does so.
Along with the focus on integers in the formalization, we will assume that I/O programs to test are written using the two primitive operations \ensuremath{\Varid{readLn}} and \ensuremath{\Varid{print}} that incorporate conversion from and to strings\footnote{This choice of primitive operations is also not meant as a real restriction.
In the actual implementation we also provide primitives that directly operate on strings.}.

\section{Current practice}
\label{sec:overview}

Consider the following verbal description of a function one might give as an exercise task to a beginning programmer:
\begin{center}
  \textit{``Add up all the numbers in a given list.''}
\end{center}
A simple Haskell solution could look like this:
\begin{hscode}\SaveRestoreHook
\column{B}{@{}>{\hspre}l<{\hspost}@{}}%
\column{3}{@{}>{\hspre}l<{\hspost}@{}}%
\column{15}{@{}>{\hspre}l<{\hspost}@{}}%
\column{E}{@{}>{\hspre}l<{\hspost}@{}}%
\>[3]{}\Varid{sum}\mathbin{::}[\mskip1.5mu \Conid{Int}\mskip1.5mu]\to \Conid{Int}{}\<[E]%
\\
\>[3]{}\Varid{sum}\;[\mskip1.5mu \mskip1.5mu]{}\<[15]%
\>[15]{}\mathrel{=}\mathrm{0}{}\<[E]%
\\
\>[3]{}\Varid{sum}\;(\Varid{x}\mathbin{:}\Varid{xs}){}\<[15]%
\>[15]{}\mathrel{=}\Varid{x}\mathbin{+}\Varid{sum}\;\Varid{xs}{}\<[E]%
\ColumnHook
\end{hscode}\resethooks
To test this solution, we could use QuickCheck properties like the following ones:
\begin{adjustedcenter}
\begin{minipage}[t]{.3\textwidth}
\begin{hscode}\SaveRestoreHook
\column{B}{@{}>{\hspre}l<{\hspost}@{}}%
\column{3}{@{}>{\hspre}l<{\hspost}@{}}%
\column{E}{@{}>{\hspre}l<{\hspost}@{}}%
\>[3]{}\Varid{propSing}\mathbin{::}\Conid{Int}\to \Conid{Bool}{}\<[E]%
\\
\>[3]{}\Varid{propSing}\mathrel{=}\lambda \Varid{x}\to \Varid{sum}\;[\mskip1.5mu \Varid{x}\mskip1.5mu]\mathrel{==}\Varid{x}{}\<[E]%
\ColumnHook
\end{hscode}\resethooks
\end{minipage}
\hfill
\begin{minipage}[t]{.65\textwidth}
\begin{hscode}\SaveRestoreHook
\column{B}{@{}>{\hspre}l<{\hspost}@{}}%
\column{3}{@{}>{\hspre}l<{\hspost}@{}}%
\column{E}{@{}>{\hspre}l<{\hspost}@{}}%
\>[3]{}\Varid{propAdd}\mathbin{::}[\mskip1.5mu \Conid{Int}\mskip1.5mu]\to [\mskip1.5mu \Conid{Int}\mskip1.5mu]\to \Conid{Bool}{}\<[E]%
\\
\>[3]{}\Varid{propAdd}\mathrel{=}\lambda \Varid{xs}\;\Varid{ys}\to \Varid{sum}\;(\Varid{xs}\plus \Varid{ys})\mathrel{==}\Varid{sum}\;\Varid{xs}\mathbin{+}\Varid{sum}\;\Varid{ys}{}\<[E]%
\ColumnHook
\end{hscode}\resethooks
\end{minipage}
\end{adjustedcenter}
Now consider another task, which might appear in a course section introducing I/O programs%
:
\begin{center}
  \textit{``Read a natural number n from \texttt{stdin}, then read n additional numbers and print the sum of those n numbers to \texttt{stdout}.''}
\end{center}
The following Haskell solution has basically the same computational content as the function further above.
But the fact that the program has to fetch its inputs on its own, and to report the computed result value back to the user, changes the overall code structure considerably.
\begin{adjustedcenter}
\begin{minipage}[t]{.3\textwidth}\begin{hscode}\SaveRestoreHook
\column{B}{@{}>{\hspre}l<{\hspost}@{}}%
\column{3}{@{}>{\hspre}l<{\hspost}@{}}%
\column{15}{@{}>{\hspre}l<{\hspost}@{}}%
\column{E}{@{}>{\hspre}l<{\hspost}@{}}%
\>[3]{}\Varid{main}\mathbin{::}\Conid{IO}\;(){}\<[E]%
\\
\>[3]{}\Varid{main}\mathrel{=}\mathbf{do}\;{}\<[15]%
\>[15]{}\Varid{n}\leftarrow \Varid{readLn}{}\<[E]%
\\
\>[15]{}\Varid{go}\;\Varid{n}\;\mathrm{0}{}\<[E]%
\ColumnHook
\end{hscode}\resethooks
\end{minipage}
\hfill
\begin{minipage}[t]{.65\textwidth}\begin{hscode}\SaveRestoreHook
\column{B}{@{}>{\hspre}l<{\hspost}@{}}%
\column{3}{@{}>{\hspre}l<{\hspost}@{}}%
\column{7}{@{}>{\hspre}l<{\hspost}@{}}%
\column{10}{@{}>{\hspre}l<{\hspost}@{}}%
\column{20}{@{}>{\hspre}l<{\hspost}@{}}%
\column{E}{@{}>{\hspre}l<{\hspost}@{}}%
\>[3]{}\Varid{go}\mathbin{::}\Conid{Int}\to \Conid{Int}\to \Conid{IO}\;(){}\<[E]%
\\
\>[3]{}\Varid{go}\;{}\<[7]%
\>[7]{}\mathrm{0}\;{}\<[10]%
\>[10]{}\Varid{res}\mathrel{=}\Varid{print}\;\Varid{res}{}\<[E]%
\\
\>[3]{}\Varid{go}\;{}\<[7]%
\>[7]{}\Varid{n}\;{}\<[10]%
\>[10]{}\Varid{res}\mathrel{=}\mathbf{do}\;{}\<[20]%
\>[20]{}\Varid{x}\leftarrow \Varid{readLn}{}\<[E]%
\\
\>[20]{}\Varid{go}\;(\Varid{n}\mathbin{-}\mathrm{1})\;(\Varid{x}\mathbin{+}\Varid{res}){}\<[E]%
\ColumnHook
\end{hscode}\resethooks
\end{minipage}
\end{adjustedcenter}
Now how do we test such a program?
How, even, can we describe more formally than in the second verbal description above what behavior is desired?

First we need to consider what it is that we want to test.
In the case of the simple \ensuremath{\Varid{sum}} function, we wanted to test the result value of the computation.
In the I/O case, we are also interested in the interaction of the program with the outside world (in what order are which values read and printed, etc.).
We therefore can no longer view such programs as just mappings from input values to output values.
Instead, programs will result in a sequence of potentially interleaved input and output actions.
We call such a sequence a trace of a program.
If we want to check whether some program exhibits a certain desired behavior, we have to check the traces it can produce.

For the above I/O task description, the set of intended traces is basically
$$ \{ \traceIn{0}\traceOut{0}\traceStop, \traceIn{1}\traceIn{v_1}\traceOut{v_1}\traceStop, \traceIn{2}\traceIn{v_1}\traceIn{v_2}\traceOut{(v_1+v_2)}\traceStop, \dots \} $$
where each $ \traceIn{} $ stands for an input action and each $ \traceOut{} $ for an output action.
If we now assume that we never supply negative numbers as input values (at least not for the first input), then the Haskell I/O program given above indeed produces exactly all, and only, traces from this set.

Following the approach presented by Swierstra and Altenkirch~\cite{swierstra2007}, corresponding tests can be automated.
First, an alternative monad is defined that represents a semantic domain for console I/O programs\footnote{Swierstra and Altenkirch use a representation %
  based on input and output of single characters.
We are currently not interested in such a fine-grained inspection and therefore always require programs to read or write whole lines.
}:
\begin{adjustedcenter}
\begin{minipage}[t]{.35\textwidth}
\begin{hscode}\SaveRestoreHook
\column{B}{@{}>{\hspre}l<{\hspost}@{}}%
\column{3}{@{}>{\hspre}l<{\hspost}@{}}%
\column{5}{@{}>{\hspre}c<{\hspost}@{}}%
\column{5E}{@{}l@{}}%
\column{8}{@{}>{\hspre}l<{\hspost}@{}}%
\column{E}{@{}>{\hspre}l<{\hspost}@{}}%
\>[3]{}\mathbf{data}\;\mathit{IO}_\mathit{rep}\;\Varid{a}{}\<[E]%
\\
\>[3]{}\hsindent{2}{}\<[5]%
\>[5]{}\mathrel{=}{}\<[5E]%
\>[8]{}\Conid{GetLine}\;(\Conid{String}\to \mathit{IO}_\mathit{rep}\;\Varid{a}){}\<[E]%
\\
\>[3]{}\hsindent{2}{}\<[5]%
\>[5]{}\mid {}\<[5E]%
\>[8]{}\Conid{PutLine}\;\Conid{String}\;(\mathit{IO}_\mathit{rep}\;\Varid{a}){}\<[E]%
\\
\>[3]{}\hsindent{2}{}\<[5]%
\>[5]{}\mid {}\<[5E]%
\>[8]{}\Conid{Return}\;\Varid{a}{}\<[E]%
\ColumnHook
\end{hscode}\resethooks
\end{minipage}
\hfill
\begin{minipage}[t]{.6\textwidth}
\begin{hscode}\SaveRestoreHook
\column{B}{@{}>{\hspre}l<{\hspost}@{}}%
\column{3}{@{}>{\hspre}l<{\hspost}@{}}%
\column{5}{@{}>{\hspre}l<{\hspost}@{}}%
\column{19}{@{}>{\hspre}l<{\hspost}@{}}%
\column{E}{@{}>{\hspre}l<{\hspost}@{}}%
\>[3]{}\mathbf{instance}\;\Conid{Monad}\;\mathit{IO}_\mathit{rep}\;\mathbf{where}{}\<[E]%
\\
\>[3]{}\hsindent{2}{}\<[5]%
\>[5]{}\Conid{GetLine}\;\Varid{f}{}\<[19]%
\>[19]{}\bind \Varid{g}\mathrel{=}\Conid{GetLine}\;(\lambda \Varid{s}\to \Varid{f}\;\Varid{s}\bind \Varid{g}){}\<[E]%
\\
\>[3]{}\hsindent{2}{}\<[5]%
\>[5]{}\Conid{PutLine}\;\Varid{s}\;\Varid{ma}{}\<[19]%
\>[19]{}\bind \Varid{g}\mathrel{=}\Conid{PutLine}\;\Varid{s}\;(\Varid{ma}\bind \Varid{g}){}\<[E]%
\\
\>[3]{}\hsindent{2}{}\<[5]%
\>[5]{}\Conid{Return}\;\Varid{a}{}\<[19]%
\>[19]{}\bind \Varid{g}\mathrel{=}\Varid{g}\;\Varid{a}{}\<[E]%
\\[\blanklineskip]%
\>[3]{}\hsindent{2}{}\<[5]%
\>[5]{}\Varid{return}\mathrel{=}\Conid{Return}{}\<[E]%
\ColumnHook
\end{hscode}\resethooks
\end{minipage}
\end{adjustedcenter}
Next, the \ensuremath{\Conid{IO}} primitives to be used are implemented for this new representation:
\begin{adjustedcenter}
\begin{minipage}[t]{.4\textwidth}
\begin{hscode}\SaveRestoreHook
\column{B}{@{}>{\hspre}l<{\hspost}@{}}%
\column{3}{@{}>{\hspre}l<{\hspost}@{}}%
\column{E}{@{}>{\hspre}l<{\hspost}@{}}%
\>[3]{}\Varid{readLn}\mathbin{::}\Conid{Read}\;\Varid{a}\Rightarrow \mathit{IO}_\mathit{rep}\;\Varid{a}{}\<[E]%
\\
\>[3]{}\Varid{readLn}\mathrel{=}\Varid{fmap}\;\Varid{read}\;(\Conid{GetLine}\;\Conid{Return}){}\<[E]%
\ColumnHook
\end{hscode}\resethooks
\end{minipage}
\hfill
\begin{minipage}[t]{.55\textwidth}
\begin{hscode}\SaveRestoreHook
\column{B}{@{}>{\hspre}l<{\hspost}@{}}%
\column{3}{@{}>{\hspre}l<{\hspost}@{}}%
\column{E}{@{}>{\hspre}l<{\hspost}@{}}%
\>[3]{}\Varid{print}\mathbin{::}\Conid{Show}\;\Varid{a}\Rightarrow \Varid{a}\to \mathit{IO}_\mathit{rep}\;(){}\<[E]%
\\
\>[3]{}\Varid{print}\;\Varid{x}\mathrel{=}\Conid{PutLine}\;(\Varid{show}\;\Varid{x})\;(\Conid{Return}\;()){}\<[E]%
\ColumnHook
\end{hscode}\resethooks
\end{minipage}
\end{adjustedcenter}
Now, any potential Haskell solution to the I/O task given further above, \ensuremath{\Varid{main}\mathrel{=}\mathbf{do}~\dots}, can not only be used at type \ensuremath{\Conid{IO}\;()}, but also at type \ensuremath{\mathit{IO}_\mathit{rep}\;()}.
Since values of that type are more inspectable than those of a normal \ensuremath{\Conid{IO}} type, we can then ``run'' \ensuremath{\Varid{main}} in a kind of simulation mode that produces an explicit trace as a data structure when given some concrete inputs:
\begin{adjustedcenter}
\begin{minipage}[t]{.55\textwidth}
\begin{hscode}\SaveRestoreHook
\column{B}{@{}>{\hspre}l<{\hspost}@{}}%
\column{3}{@{}>{\hspre}l<{\hspost}@{}}%
\column{25}{@{}>{\hspre}l<{\hspost}@{}}%
\column{33}{@{}>{\hspre}l<{\hspost}@{}}%
\column{E}{@{}>{\hspre}l<{\hspost}@{}}%
\>[3]{}\mathit{run}_\mathit{rep}\mathbin{::}\mathit{IO}_\mathit{rep}\;()\to [\mskip1.5mu \Conid{String}\mskip1.5mu]\to \Conid{Trace}{}\<[E]%
\\
\>[3]{}\mathit{run}_\mathit{rep}\;(\Conid{GetLine}\;\Varid{f})\;{}\<[25]%
\>[25]{}(\Varid{x}\mathbin{:}\Varid{xs}){}\<[33]%
\>[33]{}\mathrel{=}\Conid{Read}\;\Varid{x}\;(\mathit{run}_\mathit{rep}\;(\Varid{f}\;\Varid{x})\;\Varid{xs}){}\<[E]%
\\
\>[3]{}\mathit{run}_\mathit{rep}\;(\Conid{PutLine}\;\Varid{s}\;\Varid{ma})\;{}\<[25]%
\>[25]{}\Varid{xs}{}\<[33]%
\>[33]{}\mathrel{=}\Conid{Write}\;\Varid{s}\;(\mathit{run}_\mathit{rep}\;\Varid{ma}\;\Varid{xs}){}\<[E]%
\\
\>[3]{}\mathit{run}_\mathit{rep}\;(\Conid{Return}\;())\;{}\<[25]%
\>[25]{}[\mskip1.5mu \mskip1.5mu]{}\<[33]%
\>[33]{}\mathrel{=}\Conid{Stop}{}\<[E]%
\ColumnHook
\end{hscode}\resethooks
\end{minipage}
\hfill
\begin{minipage}[t]{.4\textwidth}
\begin{hscode}\SaveRestoreHook
\column{B}{@{}>{\hspre}l<{\hspost}@{}}%
\column{3}{@{}>{\hspre}l<{\hspost}@{}}%
\column{5}{@{}>{\hspre}c<{\hspost}@{}}%
\column{5E}{@{}l@{}}%
\column{8}{@{}>{\hspre}l<{\hspost}@{}}%
\column{E}{@{}>{\hspre}l<{\hspost}@{}}%
\>[3]{}\mathbf{data}\;\Conid{Trace}{}\<[E]%
\\
\>[3]{}\hsindent{2}{}\<[5]%
\>[5]{}\mathrel{=}{}\<[5E]%
\>[8]{}\Conid{Read}\;\Conid{String}\;\Conid{Trace}{}\<[E]%
\\
\>[3]{}\hsindent{2}{}\<[5]%
\>[5]{}\mid {}\<[5E]%
\>[8]{}\Conid{Write}\;\Conid{String}\;\Conid{Trace}{}\<[E]%
\\
\>[3]{}\hsindent{2}{}\<[5]%
\>[5]{}\mid {}\<[5E]%
\>[8]{}\Conid{Stop}{}\<[E]%
\ColumnHook
\end{hscode}\resethooks
\end{minipage}
\end{adjustedcenter}
If we now assume existence of a predicate \ensuremath{\Varid{checkCorrectness}\mathbin{::}\Conid{Trace}\to \Conid{Bool}} and of a random generator \ensuremath{\Varid{validInputs}\mathbin{::}\Conid{Gen}\;[\mskip1.5mu \Conid{String}\mskip1.5mu]}, then we can use QuickCheck again to automatically test whether a program has the intended behavior.
But writing such a generator and in particular the predicate is generally not as straightforward as one might hope.
Additionally, requiring feedback in case the predicate returns \ensuremath{\Conid{False}} adds significant extra complexity.

Our specification language and surrounding tooling solve exactly these problems.
We can state what behavior we want to see, and automatically get the relevant components required for testing.

\newcommand{\blank}{\textbf{?}}
\newcommand{\bigblank}{\textbf{???}}

\section{Specifications}\label{sec:design}
The main goal of the specification language is to describe the behavior that a correct solution for some task should have.
We want such descriptions to be concise, intuitive and easily adaptable toward new tasks.
The design currently does not include any abstraction facilities or try to achieve compositionality since at the moment we focus only on specifying small scale exercise tasks.
For the same reason, we do not care to capture all possible console I/O programs, or rather their behavior.
A lot of behavior is uninteresting or ill-suited for exercise tasks.
For example, we want behavior that necessarily requires interaction, so that students actually have to make use of the I/O-primitives to solve the task.
Section~\ref{sec:restrictions} will go into more detail regarding this aspect.
To motivate the design of our small DSL, we will first go through the example task from above step by step and see what constructs are necessary to describe the intended behavior formally (Section~\ref{sec:dsl}).
We also explain how a specification is to be interpreted intuitively in terms of execution traces (Section~\ref{sec:valid-runs}).

\subsection{Describing behavior}\label{sec:dsl}

Recall the second task description from the previous section:
\begin{center}
  \textit{``Read a natural number n from \texttt{stdin}, then read n additional numbers and print the sum of those n numbers to \texttt{stdout}.''}
\end{center}

Since we want to speak about interactive behavior, we first need notations for input and output primitives.
We use square brackets to describe such atomic actions and distinguish inputs from outputs via a triangle arrow into something or out of something.
This gives us the following basic skeleton for our specification:
$$ \readInput{\blank}{}\: \bigblank\: \writeOutputSimple{\blank} $$
We use (silent) concatenation to glue several shorter specifications together.
The above skeleton already encodes that we first read something and at some later point should print something back.

Next, in order to relate inputs and outputs, we need variables to reference read values at later points and functions to express computations over the values referenced by those variables:
$$ \readInput{n}{} \: \bigblank \: \writeOutputSimple{\mathit{sum}(\blank)} $$
Right now it is not clear what the argument to $ \mathit{sum} $ should be, but we will fill it in shortly.

The middle part of our example specification should correspond to the reading-in of the $ n $ numbers we want to sum.
Since $ n $ is determined by the first read value, we do not know up front (before a program runs) how many values we need to read overall.
Therefore we need some mechanism for flexible iteration (rather than just some fixed times concatenation of sub-specifications).
We mark the part of a specification we would like to iterate with {\loopArr} and introduce a marker {\loopExit} to indicate where/when the iteration process should finish:
$$ \readInput{n}{} (\bigblank\: \loopExit)^\loopArr \writeOutputSimple{\mathit{sum}(\blank)} $$
Now the middle part is repeated until the exit marker is hit.
However, up to now we have no way to skip over certain parts of a specification or to choose between alternatives based on some condition.
In order for our iteration process to not always terminate after the first round, we need to introduce a branching construct:
$$ \readInput{n}{} (\blank \restCh{\blank} \loopExit)^\loopArr \writeOutputSimple{\mathit{sum}(\blank)} $$
Now we can fill in a condition that only when satisfied gives control to the right branch, leading in our case to the termination of the iteration process.
Otherwise the left branch will be used.

We can use this new construct to repeatedly read in a value:
$$ \readInput{n}{} (\readInput{x}{} \restCh{\blank} \loopExit)^\loopArr \writeOutputSimple{\mathit{sum}(\blank)} $$
But now we have a problem, or actually two.
In each round the old value we ``assigned'' to $ x $ previously is lost, and we have no way of knowing when to stop.
The key feature of our DSL that helps solve both issues is the fact that variables do not just store a current value like in most programming languages.
Variables instead hold lists of all values assigned to them in chronological order.
There are then two different ways to access a variable, either as the traditional current value, denoted via the subscript $C$ (current), or as the list of all values read into that variable so far, denoted with the subscript $A$ (all).
This gives us the expressive power to not only construct the missing branching condition but now also fill in the missing argument to the summation:
$$ \readInput{n}{} (\readInput{x}{} \restCh{\mathit{len}(x_A)=n_C} \loopExit)^\loopArr \writeOutputSimple{\mathit{sum}(x_A)} $$
One thing the verbal description states that is not yet present in the DSL expression is the fact that the first number should not be negative.
This kind of restriction (in a task) is often useful when we do not care about ill-formed or otherwise undesirable inputs, especially in an educational setting where we usually introduce new concepts one step at a time.
That is, in the beginning of a course we might not want students to, for example, have to worry about checking inputs for correctness.
But later on we might explicitly require them to do so.
Our specification language therefore provides the necessary flexibility to go both ways.
Each occurrence of the primitive for reading has to be annotated with the set of values we expect there (and the way in which the specification will then be used determines whose job it is to take care of those expectations, the students' and/or the tester's\footnote{In our current setting we only ever present well-formed inputs to the students' programs.}):
$$ \readInput{n}{\nat} (\readInput{x}{\integer} \restCh{\mathit{len}(x_A)=n_C} \loopExit)^\loopArr \writeOutputSimple{\mathit{sum}(x_A)} $$
The specification we have arrived at now (and which is essentially, up to a minuscule syntactic difference, already a valid expression in our DSL) is quite rigid, as there is no flexibility with regard to the interaction allowed.
Continuing our example, one might want to allow the programs to have some extra behavior that does not really influence the core functionality.
For example, we could modify the previous task description as follows:
\begin{center}
  \textit{``Read a natural number n from \texttt{stdin}, then read n additional numbers and print the sum of those n numbers to \texttt{stdout}.
    \textbf{Additionally, when the program is still expecting at least one additional summand, it might print how many more summands it is expecting, before reading in the next input.}''}
\end{center}
We encode such optional behavior directly inside the output primitive.
That is, instead of giving a single term to describe what we expect as output, we use a set of possible terms.
This set might contain the ``empty'' term $ \varepsilon $ representing no output and thereby optionality:
$$ \readInput{n}{\nat} \big( (\writeOutput{\varepsilon, n_C - \mathit{len}(x_A)}\readInput{x}{\integer} ) \restCh{\mathit{len}(x_A)=n_C} \loopExit\big)^\loopArr \writeOutput{\mathit{sum}(x_A)} $$
While this way of expressing optionality can look a bit cumbersome compared to, for example, simply flagging an output as optional via a dedicated construct, it is far more expressive since the set we can give there is rather arbitrary.
For example, we could allow the programs, for whatever reason, to output exactly any multiple of the result of some value computation%
\footnote{This expressiveness really pays off when we move to outputting arbitrary strings, since we can then specify that we allow any output string as long as it contains the required result somewhere.}.

At first glance it might seem overly complicated or restrictive that we only introduce this specific kind of non-determinism, in outputs, and not, for example, a general non-deterministic choice operator.
But such a more general operator would allow us to write specifications that represent statements like ``Fulfill either task A or task B''.
If we now make such a specification part of an iteration expression, a different task can be chosen to be fulfilled in each round.
Abstractly this is fine, but since programs (in the language in which students write their submissions) are not actually capable of true non-deterministic choice, we cannot have a program that really behaves that way.
Therefore such a choice operator introduces not only the form of optionality we do want to encode; instead it enables also a form of meta statement that we think should not be part of the specification language itself.

Note that this does not mean that specifications cannot require completely different behavior depending on some input.
For example, we can write specifications of the form $ \readInput{x}{\integer} ( s_1 \restCh{p(x_C)} s_2 ) $.
But since $p(x_C)$ is deterministically defined once $x_C$ is known, there is no non-determinism involved here.
Combining this kind of deterministic branching with the possibility to have an empty specification, which we denote by $\nop$, we can write specifications like $ \nop \restCh{p(x_C)} s $, which only requires $s$ to be fulfilled if $p(x_C)$ evaluates to $\mathit{True}$.

\subsection{Valid program runs}\label{sec:valid-runs}

Consider now the following trace we might get from a program: $ \traceIn{2}\traceIn{5}\traceIn{3}\traceOut{8}\traceStop$.
The program first reads in the numbers 2, 5 and 3, then prints 8 and stops.
Does this trace match the specification developed above, i.e., could a program fulfilling the specification have such a run?
If not, we have just found evidence that the program under consideration does not fulfill the specification.
We can check the validity of the trace by going from left to right (and possibly in loops) through the specification and seeing if the trace actions match the required actions, while keeping track of the contents of variables.

Starting with $ \traceIn{2} $, we compare it to $ \readInput{n}{\nat} $.
Since both are input actions and moreover $2$ is a natural number, as required, we continue by checking the remaining trace against the rest of the specification.

Next we have to check the iteration.
To do this, we first check the trace against the iteration body while remembering the context in which the iteration occurred, i.e., the specification following it and the iteration body we might have to repeat.
When we hit the end (but not exit marker) of the body, that is, we did not encounter an {\loopExit}, we just check the remaining trace against the iteration body again.
When we do encounter an exit marker, we continue by checking the remaining trace against the specification following the whole iteration.

For our current case, the iteration's body is $(\writeOutput{\varepsilon, n_C - \mathit{len}(x_A)}\readInput{x}{\integer}) \restCh{\mathit{len}(x_A)=n_C} \loopExit $.
So we have to check $ \traceIn{5}\traceIn{3}\traceOut{8}\traceStop $ against that.
We first evaluate the branching condition to determine which sub-specification we have to match against.
Since $ x_A $ has length $0$ at the moment, we choose the left branch, which means we have to check $ \traceIn{5} $ against $ \writeOutput{\varepsilon, n_C - \mathit{len}(x_A)} $.
The trace action here is an input, but the specification calls for an output action.
However, since $ \varepsilon $ is contained in the set of possible outputs, this is not problematic.
After all, we can simply skip this output step, hoping we then find a match for the trace action.
And indeed the next action required by the specification is $ \readInput{x}{\integer} $, which matches $ \traceIn{5} $ and results in $5$ being assigned to $ x $ (actually, to be assigned to $x_C$ and appended to $x_A$).
Since we have no specification left to check locally, but are inside an iteration, we again check against the whole iteration body.
This results in $3$ also being read into $x$, which now conceptually holds the list $ [5,3] $ (as $x_A$, with $x_C$ being the $3$ from the end of that list).
Therefore, in the next round the branching condition evaluates to $\mathit{True}$, thus ending the loop due to the occurrence of $\loopExit$ found to the right of the branching construct.

All that is left now is to check $ \traceOut{8}\traceStop $ against $ \writeOutput{\mathit{sum}(x_A)} $.
Since we have $ \mathit{sum}(x_A) = \mathit{sum}([5,3]) = 8 $, this check is positive, also taking into account that $\traceStop$ matches the empty specification.
Overall, we can conclude that the trace $ \traceIn{2}\traceIn{5}\traceIn{3}\traceOut{8}\traceStop $ is a valid program run for the specification.

An important feature of our specification language is that we also can essentially reverse this procedure, then looking for a (random) trace that will match the specification (instead of starting from a given trace).
For the example specification above, this could yield $ \traceIn{3}\traceOut{\{\varepsilon, 3\}}\traceIn{-1}\traceOut{\{\varepsilon, 2\}}\traceIn{7}\traceOut{\{\varepsilon, 1\}}\traceIn{4}\traceOut{\{10\}}\traceStop $ as one possible trace form, where the values of inputs are random elements of the expected types and the output values are the results of evaluating all output possibilities at the respective points.
Such generalized traces can then be used to test programs by checking if a program's trace for the same input sequence is covered by the ``specification trace''.
For example, if a program produces the trace $ \traceIn{3}\traceOut{4}\traceIn{-1}\traceOut{2}\traceIn{7}\traceOut{1}\traceIn{4}\traceOut{10}\traceStop $, we see that this is not a valid trace since the first output action does not actually allow the printing of $4$.
If we do this for enough input sequences of random traces derived from the specification, we either find a counterexample or gain reasonable confidence in the correctness of a program (submission).

\subsection{Restrictions on expressiveness}\label{sec:restrictions}

As we already hinted at earlier, the expressiveness of the specification language is restricted at several points.
That rules out specifications of certain kinds of behavior, for good or bad.
Most notably, we deliberately ruled out general non-determinism, as already explained in Section~\ref{sec:dsl}.
On the one hand, such restrictions keep the syntax and semantics of the specification language simple and have the potential to enable additional reasoning about specifications.
On the other hand, they are motivated by our interpretation of a single specification as describing exactly one pattern of behavior, and one we actually consider useful in our educational setting at that.
Concerning the latter point, we for example would like the specified pattern to enforce actual interactivity.
That is, at its core the behavior should rely on a (somewhat alternating) sequence of reads and writes and should not be expressible in a different way.
Consider, for example, the following Haskell program:
\begin{adjustedcenter}
\begin{minipage}[t]{.3\textwidth}
\begin{hscode}\SaveRestoreHook
\column{B}{@{}>{\hspre}l<{\hspost}@{}}%
\column{3}{@{}>{\hspre}l<{\hspost}@{}}%
\column{14}{@{}>{\hspre}l<{\hspost}@{}}%
\column{E}{@{}>{\hspre}l<{\hspost}@{}}%
\>[3]{}\Varid{main}\mathbin{::}\Conid{IO}\;(){}\<[E]%
\\
\>[3]{}\Varid{main}\mathrel{=}\mathbf{do}\;{}\<[14]%
\>[14]{}\Varid{n}\leftarrow \Varid{readLn}{}\<[E]%
\\
\>[14]{}\Varid{loop}\;\Varid{n}{}\<[E]%
\ColumnHook
\end{hscode}\resethooks
\end{minipage}
\hfill
\begin{minipage}[t]{.65\textwidth}
\begin{hscode}\SaveRestoreHook
\column{B}{@{}>{\hspre}l<{\hspost}@{}}%
\column{3}{@{}>{\hspre}l<{\hspost}@{}}%
\column{20}{@{}>{\hspre}l<{\hspost}@{}}%
\column{E}{@{}>{\hspre}l<{\hspost}@{}}%
\>[3]{}\Varid{loop}\mathbin{::}\Conid{Int}\to \Conid{IO}\;(){}\<[E]%
\\
\>[3]{}\Varid{loop}\;\Varid{n}\mid \Varid{n}\leq \mathrm{0}{}\<[20]%
\>[20]{}\mathrel{=}\Varid{return}\;(){}\<[E]%
\\
\>[3]{}\Varid{loop}\;\Varid{n}{}\<[20]%
\>[20]{}\mathrel{=}\Varid{print}\;\Varid{n}\sequ \Varid{loop}\;(\Varid{n}\mathbin{-}\mathrm{1}){}\<[E]%
\ColumnHook
\end{hscode}\resethooks
\end{minipage}
\end{adjustedcenter}
A specification corresponding to this program is not expressible in our DSL (reading and writing integer values), and that was a design goal.
The non-expressibility is due to the facts that in our specifications an iteration process can only end based on some predicate over the global variable state (contents of variables, their history) and that only inputs can alter this state, leaving the above kind of ``output-driven loops'' impossible to encode.
According to our motivation,
this restriction is a good thing.
We only want inherently interactive behavior to be expressible, whereas the above program can be rewritten as
\begin{adjustedcenter}
\begin{minipage}[t]{.3\textwidth}
\begin{hscode}\SaveRestoreHook
\column{B}{@{}>{\hspre}l<{\hspost}@{}}%
\column{3}{@{}>{\hspre}l<{\hspost}@{}}%
\column{14}{@{}>{\hspre}l<{\hspost}@{}}%
\column{E}{@{}>{\hspre}l<{\hspost}@{}}%
\>[3]{}\Varid{main}\mathbin{::}\Conid{IO}\;(){}\<[E]%
\\
\>[3]{}\Varid{main}\mathrel{=}\mathbf{do}\;{}\<[14]%
\>[14]{}\Varid{n}\leftarrow \Varid{readLn}{}\<[E]%
\\
\>[14]{}\Varid{print}\;(\Varid{loop}\;\Varid{n}){}\<[E]%
\ColumnHook
\end{hscode}\resethooks
\end{minipage}
\hfill
\begin{minipage}[t]{.65\textwidth}
\begin{hscode}\SaveRestoreHook
\column{B}{@{}>{\hspre}l<{\hspost}@{}}%
\column{3}{@{}>{\hspre}l<{\hspost}@{}}%
\column{20}{@{}>{\hspre}l<{\hspost}@{}}%
\column{E}{@{}>{\hspre}l<{\hspost}@{}}%
\>[3]{}\Varid{loop}\mathbin{::}\Conid{Int}\to \Conid{String}{}\<[E]%
\\
\>[3]{}\Varid{loop}\;\Varid{n}\mid \Varid{n}\leq \mathrm{0}{}\<[20]%
\>[20]{}\mathrel{=}\text{\ttfamily \char34 \char34}{}\<[E]%
\\
\>[3]{}\Varid{loop}\;\Varid{n}{}\<[20]%
\>[20]{}\mathrel{=}\Varid{show}\;\Varid{n}\plus \text{\ttfamily \char34 \char92 n\char34}\plus \Varid{loop}\;(\Varid{n}\mathbin{-}\mathrm{1}){}\<[E]%
\ColumnHook
\end{hscode}\resethooks
\end{minipage}
\end{adjustedcenter}
with exactly one input action at the beginning, then all computation happening in a non-I/O loop, and exactly one output action at the end, which overall is not an attractive teaching example when we actually want to cover interactive I/O in Haskell and how programs must be structured to organize sequences of input and output actions in interesting ways.

If we for a moment would lift our restriction to just use integers as inputs and outputs, we could write a specification like $ \readInput{n}{\integer}\writeOutput{\mathit{loop}(n)} $ for such behavior, with the second version of \ensuremath{\Varid{loop}} above.
From this it is immediately clear that the interactive core of the program/task here is almost trivial, so we do not want it.
Put differently, we wanted to make sure that there are as few as possible ways in our DSL to encode essentially non-interactive computations in only seemingly interactive guise.
Note that even if we do indeed allow strings for input and output, as we do for practical usage in our course, we can still prevent creation of such ``boring tasks'' via the DSL by controlling which functions are allowed in terms for conditions and outputs, for example preventing something like $\mathit{loop}$ from appearing there as it does in the hypothetical specification $\readInput{n}{\integer}\writeOutput{\mathit{loop}(n)}$.
We will see in the next section that this is encoded in the definition of valid terms by parameterizing it over some set of available functions.

\section{Syntax}
\label{sec:syntax}

Figure~\ref{fig:syntax} gives the full syntax of our language by defining the set $ \mathit{Spec} $ of all specifications as well as the term language used for the description of output values and branching conditions.
We distinguish different subsets of the set of all terms by a subscript indicating the type of value a term evaluates to.
For example, $ T_\mathbb{\integer} $ denotes the set of all terms that evaluate to an integer and $ T_\mathbb{B} $ the set of terms evaluating to a Boolean value.
We write $ [\integer] $ instead of $ \integer^* $ for sequences of integers here, emphasizing that we are dealing with list values as opposed to words over integers.
\begin{figure}[h]
  \def\defaultHypSeparation{\hskip0pt}
  \vspace{2ex}
  \begin{center}
  \begin{tabular}{cc}
    \AxiomC{$ \tau \subseteq \integer $}
    \AxiomC{$ x \in \mathit{Var} $}
    \RightLabel{(Read)}
    \BinaryInfC{$ \readInput{x}{\tau} \in \mathit{Spec} $}
    \DisplayProof
    &
    \AxiomC{$ s_1 \in \mathit{Spec} $}
    \AxiomC{$ s_2 \in \mathit{Spec} $}
    \RightLabel{(Seq)}
    \BinaryInfC{$ s_1 \cdot s_2  \in \mathit{Spec} $}
    \DisplayProof
    \\[4ex]
    \AxiomC{$ \Theta \subseteq (T_\integer \cup \{\varepsilon\}), \Theta \setminus \{ \varepsilon \} \neq \emptyset $}
    \RightLabel{(Write)}
    \UnaryInfC{$ \writeOutputSimple{\Theta} \in \mathit{Spec} $}
    \DisplayProof
    &
    \AxiomC{$ s_1 \in \mathit{Spec} $}
    \AxiomC{$ s_2 \in \mathit{Spec} $}
    \AxiomC{$ c \in T_\mathbb{B} $}
    \RightLabel{(Branch)}
    \TrinaryInfC{$ s_1 \restCh{c} s_2  \in \mathit{Spec} $}
    \DisplayProof
  \end{tabular}\\[2.5ex]
    \AxiomC{$ s \in \mathit{Spec} $}
    \RightLabel{(Till-$ \loopExit $)}
    \UnaryInfC{$ s^\loopArr \in \mathit{Spec} $}
    \DisplayProof
    \qquad
    \AxiomC{\vphantom{S}}
    \RightLabel{(LoopExit)}
    \UnaryInfC{$ \loopExit \in \mathit{Spec} $}
    \DisplayProof
    \qquad
    \AxiomC{\vphantom{S}}
    \RightLabel{(Nop)}
    \UnaryInfC{$ \mathbf{0} \in \mathit{Spec} $}
    \DisplayProof
  \\[2ex]
  \dotfill
  \\[2ex]
    \def\defaultHypSeparation{\hskip0pt}
    \AxiomC{$ x \in \mathit{Var} $}
    \RightLabel{(Current)}
    \UnaryInfC{$ x_C \in T_\integer $}
    \DisplayProof
    \qquad \qquad
    \AxiomC{$ x \in \mathit{Var} $}
    \RightLabel{(All)}
    \UnaryInfC{$ x_A \in T_{[\integer]} $}
    \DisplayProof
    \\[2ex]
    \AxiomC{$ f : D_1\! \times\! \dots\! \times\! D_n \rightarrow D $}
    \AxiomC{$ t_1 \in T_{D_1}, \dots, t_n \in T_{D_n}  $}
    \AxiomC{$ f \in \mathit{Func}$}
    \RightLabel{(Function)}
    \TrinaryInfC{$ f( t_1, \dots, t_n) \in T_D $}
    \DisplayProof
  \end{center}
  \caption{\label{fig:syntax} Syntax of specifications (top) and terms (bottom)}
\end{figure}

With the exception of (Write), the rules are straightforward; there, we require that the set of possible output values contains at least one real term.
That is, we deliberately rule out actions of the forms $ \writeOutput{} $ and $ \writeOutput{\varepsilon} $.
Giving an empty set of terms would always result in an unsatisfiable specification; giving a singleton set containing $ \varepsilon $ would be equivalent to $ \mathbf{0} $.

For terms we restrict ourselves to %
a not further specified set $\mathit{Func}$ of functions and the elements of some variable set $ \mathit{Var} $ used in the specification, or more precisely, the different access variants of those variables.
In principle we could choose any set of functions we want, as long as evaluation of terms is well-defined.
Making $\mathit{Func}$ itself a parameter of the specification language is useful if one wants to enforce some conditions to guarantee certain properties of specifications.
We could, for example, choose $\mathit{Func}$ to be the set of all total functions if we want some guarantees on termination.
Or if we are interested in automatically generating random specifications for exercise tasks, we can control to some extent what kind of tasks are generated by choosing different sets for $\mathit{Func}$.

We make the following assumptions regarding the structure and semantic well-formedness of specifications:
\begin{itemize}[noitemsep]
  \item
    A variable $ x_C $ does not occur in a term before $ x $ occurred in an input action, since this would make the evaluation of that term fail.
    A corresponding issue does not exist for $ x_A $ since we can define it to initially evaluate to the empty list.
  \item
    Every loop eventually reaches an occurrence of {\loopExit} (given the right sequence of input values).
    If we are not interested in actual termination, we can alternatively loosen this so that every {\loopArr} just ``binds'' an occurrence of {\loopExit}, i.e., an exit marker is present but we do not analyze the branching conditions to reach it.

    The purpose of this restriction is to let specifications only ever describe finite behavior.
    In practice, this condition does not necessarily have to be checked.
    Since we cannot prevent students from submitting programs with infinite loops, we usually work with timeouts.
    When we, for example, test a model solution against an accidentally non-terminating specification during our task development activities, these very same timeouts will ``discover'' this fact.
    It is, therefore, unlikely that we will pose tasks based on such ill-formed specifications.
\end{itemize}
Additionally, sequential composition of specifications is defined to be associative, i.e., $ s_1 \cdot (s_2 \cdot s_3) = (s_1 \cdot s_2 )\cdot s_3$, therefore we can just write $ s_1 \cdot s_2 \cdot s_3 $ instead, or indeed $ s_1 \: s_2 \: s_3 $.
Also, $\mathbf{0}$ is the neutral element of sequential composition, meaning $\mathbf{0} \cdot s = s = s \cdot \mathbf{0} $.
Moreover, we define sequential composition to have higher precedence than branching and {\loopArr} to have higher precedence than sequential composition, i.e., $ s_1 \cdot s_2 \restCh{c} s_3 = (s_1 \cdot s_2) \restCh{c} s_3$ and $ s_1 \cdot s_2 ^ \loopArr = s_1 \cdot (s_2 ^ \loopArr) $.

Also note that we have no real notion of variable scope in our language.
Every variable is global and changes to it will be visible at every point in time after that change occurred.

\section{Semantics}
\label{sec:semantics}

Recall that we want to analyze I/O programs by simulating program runs and then inspecting the resulting traces.
In Section~\ref{sec:overview} we gave a data type for representing such traces.
However, since we restrict ourselves to I/O programs that only read and write integer values in the formalization, we will use here the more restrictive and more compact version of traces already informally introduced in Section~\ref{sec:valid-runs}.

Therefore a trace is a sequence of values $ v_i \in \mathbb{Z} $ marked either as input, denoted $ \traceIn{v_i} $, or as output, denoted $ \traceOut{v_i} $.
Each trace ends with the element $ \traceStop $ indicating the end of execution.
We use $ \mathit{Tr} $ to denote the set of all traces (regardless of a certain program or specification).

\subsection{Matching traces and specifications}

In order to determine whether a given trace is valid for a given specification, we define a function $\accept$ such that $\accept(s,\_)(t,\_) = \mathit{True}$ exactly if a given trace $t\in\mathit{Tr}$ exhibits behavior specified by $s\in\mathit{Spec}$.
Figure \ref{fig:semantics} gives the detailed definition of this function.
Other than the trace and the specification that the trace is checked against, the function also takes a variable environment $ \Delta $ %
and a continuation function $ k $ as additional inputs.
The continuation $ k $ takes care of managing the current iteration context, as informally described when discussing how to check an iteration in Section~\ref{sec:valid-runs}.
It encodes how to proceed if we $\mathtt{Exit}$ from the current context to an outer one or if we just $\mathtt{End}$ a round inside the current context and continue with another round of the iteration.%

\newcommand{\cond}[1]{\text{ , if }#1}
\newcommand{\otherwise}{\text{ , otherwise}}

\begin{figure}[t]
\begin{subequations}
\begin{align}
  \accept(\readInput{x}{\tau}\cdot s',k)(t, \Delta) & = \begin{cases} \label{eq:read}
    \accept(s',k)(t', \mathit{store}(x,v,\Delta)) &\cond{t = \traceIn{v}\traceVar{t'} \wedge v \in \tau}\\
    \mathit{False} &\otherwise
  \end{cases} \\
  \accept(\writeOutputSimple{\Theta} \cdot s',k)(t, \Delta) & = \begin{cases}
    \accept(\writeOutputSimple{(\Theta \setminus \{\varepsilon\})} \cdot s',k)(t, \Delta) &\cond{\varepsilon \in \Theta} \\
      \quad \vee \accept(s',k)(t, \Delta) \\
    \accept(s',k)(t', \Delta) &\cond{\varepsilon \notin \Theta \wedge t = \traceOut{v}\traceVar{t'}} \\ &\phantom{, if }\wedge v \in \mathit{eval}(\Theta,\Delta)\\
    \mathit{False} &\otherwise
  \end{cases} \\
  \accept((s_1 \restCh{c} s_2) \cdot s',k)(t, \Delta) & = \begin{cases}
    \accept(s_2 \cdot s',k)(t,\Delta) \cond{\mathit{eval}(c,\Delta) = \mathit{True}}\\
    \accept(s_1 \cdot s',k)(t,\Delta) \otherwise
  \end{cases} \label{eq:branch} \\
  \accept(s^\loopArr \cdot s',k)(t, \Delta) & = \accept(s,k')(t,\Delta)\\
    & \phantom{= \accept} \text{ with } k'(\mathit{cont}) = \smash{ \begin{cases} \nonumber
    \accept(s, k') \cond{\mathit{cont} = \mathtt{End}}\\
    \accept(s', k) \cond{\mathit{cont} = \mathtt{Exit}}
  \end{cases}} \\
  \accept(\loopExit \cdot s',k)(t,\Delta) & = k(\mathtt{Exit})(t,\Delta)  \label{eq:exit}\\
  \accept(\nop,k)(t,\Delta) & = k(\mathtt{End})(t,\Delta) \label{eq:end}\\
  k_I(\mathit{cont})(t, \Delta) & = \begin{cases}
    \mathit{True} &\cond{\mathit{cont} = \mathtt{End} \wedge t = \traceStop}\\
    \mathit{False} &\cond{\mathit{cont} = \mathtt{End} \wedge t \neq \traceStop}\\
    \mathtt{error} &\cond{\mathit{cont} = \mathtt{Exit}}
  \end{cases} \nonumber
\end{align}
\end{subequations}
\caption{\label{fig:semantics}Trace acceptance}
\end{figure}
The functions $ \mathit{eval} $ and $ \mathit{store} $ evaluate terms and store values in the environment, respectively.
Their definitions are straightforward and are therefore omitted here.
We write $ \mathit{eval}(\Theta,\Delta) $ for evaluating, under $ \Delta $, every term in a set $ \Theta $.

At its core, $\accept$ traverses a specification from left to right, consuming matching trace elements and updating variables along the way.
If we are left with exactly the empty specification or if we encounter an exit marker of some iteration, we call $k$ with the appropriate argument to indicate whether we want to continue the iteration process or exit from it, and pass the remaining trace and current variable environment along.
Note that this also covers the case where we completely consumed the specification in the outermost context, i.e., the initial one.
If the trace is then also fully consumed already, the acceptance match is successful.
The initial continuation $k_I$ is defined such that in the case of $\mathtt{End}$ it performs exactly this check (see Figure~\ref{fig:semantics}) and therefore finishes the computation.

It is important to note that the cases (\ref{eq:read}) to (\ref{eq:exit}) are all defined with the associativity of sequential composition in mind.
Therefore we do not have a case for $ \accept((s \cdot s')\cdot s'', k)(t,\Delta) $ since this can always be rewritten as $ \accept(s \cdot (s' \cdot s''), k)(t,\Delta) $.
Moreover, the fact that $\nop$ is the neutral element of sequential composition means that the rules can also be applied to specifications like $\readInput{x}{\tau}$ by expanding them to $\readInput{x}{\tau} \cdot \nop$.
By the same reasoning, we do not need an explicit rule like $\accept(\nop \cdot s',k)(t, \Delta) = \accept(s',k)(t, \Delta)$.
Additionally, (\ref{eq:end}) is only applicable if the specification is exactly $\nop$.
Otherwise this would allow for the initiation of another round of iteration at any point in the specification, since for example $s \cdot s'$ can always be rewritten as $s \cdot \nop \cdot s'$.
Note that for $\loopExit$ in (\ref{eq:exit}) the situation is different.
Even though it might seem strange at first, since one would probably never write a specification containing ``dead code'' $s'$ like this, cases exist where a specification of the form $ \loopExit \cdot s'$ does arise.
Consider, for example, $ ((s_1 \restCh{c_1} (s_2 \restCh{c_2} \loopExit)) \cdot s_3)^\loopArr $, which is a perfectly reasonable specification to write.
Now in order to leave the loop via that $\loopExit$, both conditions must evaluate to $\mathit{True}$, and by (\ref{eq:branch}) we are now left with matching against $\loopExit \cdot s_3$.
So in order to correctly handle such specifications, (\ref{eq:exit}) needs to discard everything following an occurrence of $\loopExit$.

\subsection{Program correctness}\label{sec:correctness}

\phantom{%
\parbox{\textwidth}{%
Using the $\accept$-function, we can formulate a notion of program (trace) correctness.
A program is considered to be an implementation of a specification $ s $ if and only if for every execution trace $ t $ that can arise from the program, it holds $ \accept(s,k_I)(t,\Delta_I) = \mathit{True} $, where $ \Delta_I $ is the initial environment containing no values, i.e., $\mathit{eval}(x_A,\Delta_I) $ evaluates to the empty list for every variable $x$ occurring in $ s $, and $k_I$ is the initial continuation as shown in Figure~\ref{fig:semantics}.
}}%
\begin{textblock}{1}(0,-1)
\Huge
\textcolor{gray}{\textbf{Elided}}
\end{textblock}

\section{Testing framework}
\label{sec:testing}
Now that we defined syntax and semantics of our DSL, how can we actually test programs against specifications?
Recall that we want to be able to automatically generate the following three components from a given specification:
\begin{itemize}[noitemsep]
  \item A generator of input sequences that respect the task's invariants.
  \item A way of checking whether a trace exhibits the desired behavior.
  \item A method of providing feedback in case the actual behavior did not match the expectations (e.g., a correct run on the respective input sequence).
\end{itemize}
With the $\accept$-function from above, we could already cross off the second item on the list.
For the generation of simple feedback, we could modify $\accept$ such that it returns additional information in case the matching is unsuccessful.
However, for more elaborate feedback, like example runs, we need another, less localized, approach.
One way to enable more general feedback is to take the equation $\accept(s,k_I)(t,\Delta_I) = \mathit{True}$, for some specification $s$, and solve for $t$.
A solution for $t$ is a valid program run.
Additionally, managing to compute such a $t$ also immediately gives us a valid input sequence for the specification, thereby meeting our remaining requirement.
We can further generalize this idea by not computing a concrete trace (choosing one possibility for each output action) but rather a general structure representing all accepted runs for a given input sequence.
Such a structure, called a generalized trace, then represents an all-encompassing test case for a certain fixed input sequence.
We will use generalized traces to build all of our three required components.

Given a mechanism to actually compute such test cases, we can test programs by repeatedly running the following steps:
\begin{enumerate}[noitemsep]
  \item generate a generalized trace $t_g$ from the given specification $s$,
  \item extract the input sequence from $t_g$,
  \item run the program under testing on that input sequence, resulting in trace $t_p$,
  \item check whether $t_p$ is one of the traces represented by $t_g$.
\end{enumerate}
Before we will look at how to compute generalized traces from specifications, we will first define what generalized traces are exactly and how they relate to ordinary traces.
Then we will give the definition of a modified version of $\accept$ that describes the complete set of generalized traces for a certain specification.
Even though this set is usually infinite, the function is constructed in a way that lets us derive a method for generating specific generalized traces.

\subsection{Generalized traces}
\label{sec:gen-traces}

Analogously to how we allowed different potential output values in a single output action in the specification language, we now allow different values in each output step of a trace.

Consider the specification $ \readInput{x}{\integer} \writeOutput{\varepsilon, x_C} $.
For arbitrary but fixed input value $ v \in \integer $, the valid traces for this specification are $ \traceIn{v}\traceStop $ and $ \traceIn{v}\traceOut{v}\traceStop $.
The ``test case for a fixed $v$'' should allow any of those two traces.
Therefore the generalized trace for this specification under a fixed input $v$ is $ \traceIn{v}\traceOut{\{ \varepsilon, v\}}\traceStop $.

Moreover, and unlike for specifications, we fuse sequences of adjacent output steps into a single output step, then containing (possibly various) sequences of values.
For example, if we had a specification like $ \readInput{x}{\integer} \writeOutput{\varepsilon, x_C} \writeOutput{\varepsilon, x_C} $, without this normalization we would end up with $ \traceIn{v}\traceOut{\{ \varepsilon, v\}}\traceOut{\{ \varepsilon, v\}}\traceStop $ as our generalized trace.
Since we cannot distinguish between the two outputs anyway, due to the black-box nature of our approach, we just combine them into a single action.
The normalized trace in this case is therefore $ \traceIn{v}\traceOut{\{ \varepsilon, v, vv \}}\traceStop $.
This trace completely covers the possible behavior of a correct program for specification $ \readInput{x}{\integer} \writeOutput{\varepsilon, x_C} \writeOutput{\varepsilon, x_C} $ and fixed input $v$.

The set of all \emph{generalized} traces in general, $ \mathit{Tr}_G $, is defined by the following rules:
\begin{center}
  \def\defaultHypSeparation{\hskip0pt}
\begin{tabular}{cc}
  \AxiomC{$ v \in \mathbb{Z} $}
  \AxiomC{$ t \in \mathit{Tr}_G $}
  \BinaryInfC{$ \traceIn{v}\traceVar{t} \in \mathit{Tr}_G $}
  \DisplayProof
  &
  \AxiomC{$ v \in \mathbb{Z} $}
  \AxiomC{$ t \in \mathit{Tr}_G $}
  \AxiomC{$ V \subseteq \mathbb{Z}^*$}
  \AxiomC{$ V \setminus \{ \varepsilon \} \neq \emptyset $}
  \QuaternaryInfC{$ \traceOut{V}\traceIn{v}\traceVar{t} \in \mathit{Tr}_G $}
  \DisplayProof
  \\[3ex]
  \AxiomC{\vphantom{$ \setminus $}}
  \UnaryInfC{$ \traceStop \in \mathit{Tr}_G $}
  \DisplayProof
  &
  \AxiomC{$ V \subseteq \mathbb{Z}^*$}
  \AxiomC{$ V \setminus \{\varepsilon\} \neq \emptyset $}
  \BinaryInfC{$ \traceOut{V}\traceStop \in \mathit{Tr}_G $}
  \DisplayProof
\end{tabular}
\end{center}
We now need a way to determine whether an ordinary trace is \emph{covered} by a generalized trace.
That is, we want to check whether the concrete run represented by the ordinary trace is contained in the set of runs the generalized trace is representing.
Since in a generalized trace consecutive outputs are fused, we first normalize ordinary traces in that respect as well if we want to compare them to generalized ones.
We normalize ordinary traces by way of the function $ \lceil \cdot \rceil : \mathit{Tr} \rightarrow \mathit{Tr}_G $ that embeds $\mathit{Tr}$ into $\mathit{Tr}_G$, the image being exactly the subset of $\mathit{Tr}_G$ with only singleton sets, of non-empty words, in output steps:
\begin{minipage}[t]{.5\textwidth}
\begin{align*}
  \lceil t \rceil_{\phantom{w}}         &= \quad \lceil t \rceil_\varepsilon\\
  \lceil \traceIn{v} \traceVar{t'}\rceil_w &= \;
    \begin{cases}
      \traceIn{v} \, \lceil t' \rceil_\varepsilon &\text{, if } w = \varepsilon\\
      \traceOut{\{w\}} \traceIn{v} \, \lceil t' \rceil_\varepsilon &\text{, otherwise}\\
    \end{cases}\\
\end{align*}
\end{minipage}
\begin{minipage}[t]{.4\textwidth}
\begin{align*}
  \lceil \traceOut{v} \traceVar{t'}\rceil_w &= \quad \lceil t'\rceil_{wv}\\
  \lceil \traceStop \rceil_w &=\;
     \begin{cases}
       \mathrlap{\traceStop}
        \hphantom{\traceOut{\{w\}} \traceIn{v} \lceil t'\rceil_\varepsilon} %
       &\text{, if } w = \varepsilon\\
       \traceOut{\{w\}} \traceStop &\text{, otherwise}\\
     \end{cases}
\end{align*}
\end{minipage}

\noindent
Now we can define the covering relation $\mathord{\prec} \subseteq \lceil \mathit{Tr} \rceil \times \mathit{Tr}_G$ as follows:
\begin{center}
\newcommand{\alignphantom}{\vphantom{$\{\}t'$}}
\def\defaultHypSeparation{\hskip0pt}
  \AxiomC{\alignphantom $ t_1 \prec t_2$}
  \UnaryInfC{\alignphantom$ \traceIn{v}\traceVar{t_1} \prec \traceIn{v}\traceVar{t_2} $}
  \DisplayProof
  ~
  \AxiomC{$ w \in V $}
  \AxiomC{\alignphantom$ t_1 \prec t_2 $}
  \BinaryInfC{\alignphantom$ \traceOut{\{w\}} \traceVar{t_1} \prec \traceOut{V}\!\traceVar{t_2} $}
  \DisplayProof
  ~
  \AxiomC{$ \varepsilon \in V $}
  \AxiomC{\alignphantom$t_1 \prec t_2 $}
  \BinaryInfC{\alignphantom$ \traceVar{t_1} \prec \traceOut{V}\!\traceVar{t_2} $}
  \DisplayProof
  ~
  \AxiomC{\alignphantom}
  \UnaryInfC{\alignphantom$ \traceStop \prec \traceStop $}
  \DisplayProof
\end{center}
Note that, due to the typing of this relation, neither the trace on the left nor that on the right of any occurrence of $\prec$ can contain directly consecutive output steps.
The definition of $\prec$ is also what allows us to generate feedback in case we found an invalid program run.
If $ \lceil t_p \rceil \not\prec t_g $, for some program trace $t_p$ and a generalized trace $t_g$, it could hypothetically be for one of three reasons.
First, it could be because the two traces, at some position, disagree on the value that is read in.
This case does not occur in our setting, since we use the input sequence of the generalized trace to construct the ordinary trace.
Therefore in our setting there are actually only two possible reasons for why a trace is not covered.
Either the structure of the traces does not line up, i.e., their first constructor differs and the respective step in $t_g$ cannot be skipped (i.e., it is not $\traceOut{\{\varepsilon,\dots\}}$), or for some output step there is a complete or partial mismatch between the expected and actual output, i.e., $w \not \in V$.
In both cases a simple message like ``\emph{Expected: $\dots$, but got: $\dots$}'' or ``\emph{The output value $\dots$ is not covered by $\dots$}'' can be generated and presented to the user, along with the input that triggered the error.
Additionally, the generalized trace can be given (in some pretty-printed form) to showcase all possible runs of the program on that particular input sequence.
Depending on the application setting it might however be useful to restrict this to just one particular example run, especially when the target specification is hidden, which might be the case in an educational setting.

\subsection{Computing generalized traces}

Equipped with the definition of generalized traces, we now need a way to actually compute them for a given specification.
The basic idea from the beginning of this section was to solve for $t$ in the equation $\accept(s,k_I)(t,\Delta_I) = \mathit{True}$.
We can do this by evaluating $\accept$ with $t$ unfixed and on demand extending the trace with the appropriate steps such that we never fall into the $\mathit{False}$ case.
To do this, we need to pick random elements of the annotated set for each input action and choose one of the possible outputs of each write action.
This will then give us a particular non-generalized trace that matches the specification.

Let us take $\readInput{n}{\nat} (\writeOutput{\varepsilon, n_C - \mathit{len}(x_A)}\readInput{x}{\integer} \restCh{\mathit{len}(x_A)=n_C} \loopExit)^\loopArr \writeOutput{\mathit{sum}(x_A)} $ as an example.
Running $\accept$ %
``in reverse'' could result, for example, in one of the following traces: $\traceIn{1}\traceIn{4}\traceOut{4}\traceStop$, $\traceIn{1}\traceOut{1}\traceIn{4}\traceOut{4}\traceStop$, $\traceIn{2}\traceIn{3}\traceIn{7}\traceOut{10}\traceStop$.
If we want to get a generalized trace instead, all we need to do is to not choose a single output action but rather extend the trace with all possible outputs.
This would, for example, result in merging the first two traces from above into a single generalized trace $\traceIn{1}\traceOut{\{\varepsilon,1\}}\traceIn{4}\traceOut{\{4\}}\traceStop$.
\begin{figure}[h]
  \begin{subequations}
  \begin{align}
    \mathit{traceSet}(\readInput{x}{\tau}\cdot s',k)(\Delta) & =
      \colorbox{gray!30}{$
      \bigcup_{v \in \tau} \{ \traceIn{v}\} \cdot \mathit{traceSet}(s',k)(\mathit{store}(x,v,\Delta))
      $} \\
    \mathit{traceSet}(\writeOutputSimple{\Theta} \cdot s',k)(\Delta) & =
      \colorbox{gray!30}{$
      \mathit{eval}(\Theta,\Delta) \odot \mathit{traceSet}(s',k)(\Delta) \label{writeCombine}
      $} \\
    \mathit{traceSet}((s_1 \restCh{c} s_2) \cdot s',k)(\Delta) & = \begin{cases}
      \mathit{traceSet}(s_2 \cdot s',k)(\Delta) \cond{\mathit{eval}(c,\Delta) = \mathit{True}}\\
      \mathit{traceSet}(s_1 \cdot s',k)(\Delta) \otherwise
    \end{cases} \\
    \mathit{traceSet}(s^\loopArr \cdot s',k)(\Delta) & = \mathit{traceSet}(s,k')(\Delta)\\
      & \phantom{= \mathit{trace}} \text{ with } k'(\mathit{cont}) = \smash{ \begin{cases} \nonumber
      \mathit{traceSet}(s, k') \cond{\mathit{cont} = \mathtt{End}}\\
      \mathit{traceSet}(s', k) \cond{\mathit{cont} = \mathtt{Exit}}
    \end{cases}} \\
    \mathit{traceSet}(\loopExit \cdot s',k)(\Delta) & = k(\mathtt{Exit})(\Delta) \\
    \mathit{traceSet}(\nop,k)(\Delta) & = k(\mathtt{End})(\Delta) \\
    k^T_I(\mathit{cont})(\Delta) & = \begin{cases}
      \colorbox{gray!30}{$ %
      \{\traceStop\} \hphantom{\mathtt{error}} \cond{\mathit{cont} = \mathtt{End}}
      $}\\
      \mathtt{\;error} \hphantom{\{\traceStop\}} \cond{\mathit{cont} = \mathtt{Exit}}
    \end{cases} \nonumber\\
    \ & \hspace{-34pt} %
      \colorbox{gray!30}{$
        V \odot T' = \bigcup_{t' \in T'}
          \begin{cases}
            \{ \traceOut{(V \cdot V')}\traceVar{t''} \} &\cond{t'= \traceOut{V'}\!\traceVar{t''}} \\
            \{ \traceOut{V}\!\traceVar{t'} \} &\otherwise
          \end{cases} \nonumber
      $}
  \end{align}
  \end{subequations}
  \caption{\label{fig:traceSet} Trace set generation (differences to Figure~\ref{fig:semantics} in gray)}
\end{figure}
If we now do also not choose individual inputs from the respective sets, we get a function that describes the set of all possible generalized traces for a given specification.
For the specification above, this set would be the following one:
\[
\begin{array}{l@{}l}
  S=\{ &\traceIn{0}\traceOut{\{0\}}\traceStop,\\
       &\traceIn{1}\traceOut{\{\varepsilon, 1\}}\traceIn{v_1}\traceOut{\{v_1\}}\traceStop,\\
       &\traceIn{2}\traceOut{\{\varepsilon, 2\}}\traceIn{v_1}\traceOut{\{\varepsilon, 1\}}\traceIn{v_2}\traceOut{\{v_1 + v_2\}}\traceStop,\\
       &\dots \mid v_1,v_2,\ldots \in \integer \}
\end{array}
\]
In Figure~\ref{fig:traceSet} the definition of this trace set generation function is given.
The most notable conceptual deviation from the $\accept$-function, apart from turning an acceptor into a set generator, is case (\ref{writeCombine}).
It avoids the case distinction from the corresponding part in Figure~\ref{fig:semantics} since we consider all possible output values of a write action and combine consecutive output values into single words.
The notation $V_1 \cdot V_2$ in the definition of the corresponding helper (and elsewhere in the figure, for traces) denotes language concatenation.
For notational simplicity, we also assume that $\mathit{eval}(\varepsilon,\Delta) = \varepsilon$.

The connection between $\accept$ and $\mathit{traceSet}$ can be given as follows:
Let $s \in \mathit{Spec}$ and $t \in \mathit{Tr}$, then $\accept(s,k_I)(t,\Delta_I) = \mathit{True} $ if and only if there exists a $ t' \in \mathit{traceSet}(s,k^T_I)(\Delta_I) $ such that $ \lceil t \rceil \prec t' $.

Additionally, with $\mathit{traceSet}$ we do not only have a generator for generalized traces but also an interpreter.
If we generate a generalized trace not by choosing inputs at random, but from a given sequence, we essentially execute a specification for that input sequence.

\section{Implementation}
\label{sec:impl}

We have started to build an EDSL for our design in Haskell and implemented the testing approach explained thus far%
\footnote{%
  A demo showcasing the system in the context of the automatic grading system we use is available at \url{https://autotool.fmi.iw.uni-due.de/tfpie19}.
  Note that the specifications used in that demo differ slightly from the ones presented here, as they additionally handle string inputs and outputs.
  The EDSL itself is available at \url{https://github.com/fmidue/IOTasks}.
}.
Within our framework, we provide a data type for describing a \ensuremath{\Conid{Specification}}, the \ensuremath{\mathit{IO}_\mathit{rep}} type from Section~\ref{sec:overview}, and a function \ensuremath{\Varid{fulfills}\mathbin{::}\mathit{IO}_\mathit{rep}\to \Conid{Specification}\to \Conid{Property}} that constructs a value of QuickCheck's \ensuremath{\Conid{Property}} type given a program and a specification.
When tested, this property will generate generalized traces for the given specification with the help of a QuickCheck generator based on the $\mathit{traceSet}$-function.
From such a trace we then extract the sequence of generated inputs and use \ensuremath{\mathit{run}_\mathit{rep}}, also shown in Section~\ref{sec:overview}, to generate a \ensuremath{\Conid{Trace}}.
Lastly, using $\prec$, we check if this result trace is covered by the initial generalized trace and generate an appropriate error/feedback message if that is not the case.
Note that we did not need to implement $\accept$ since, as already stated in the previous section, $\mathit{traceSet}$ together with $\prec$ can be used instead.

If we now, for example, take the specification \smash{$ \readInput{n}{\nat} (\readInput{x}{\integer} \restCh{\mathit{len}(x_A)=n_C} \loopExit)^\loopArr \writeOutput{\mathit{sum}(x_A)} $}, we can express it in the EDSL like so:
\begin{hscode}\SaveRestoreHook
\column{B}{@{}>{\hspre}l<{\hspost}@{}}%
\column{3}{@{}>{\hspre}l<{\hspost}@{}}%
\column{11}{@{}>{\hspre}l<{\hspost}@{}}%
\column{14}{@{}>{\hspre}l<{\hspost}@{}}%
\column{E}{@{}>{\hspre}l<{\hspost}@{}}%
\>[3]{}\Varid{spec}\mathbin{::}\Conid{Specification}{}\<[E]%
\\
\>[3]{}\Varid{spec}\mathrel{=}{}\<[11]%
\>[11]{}\Varid{readInput}\;\text{\ttfamily \char34 n\char34}\;\Varid{nats}{}\<[E]%
\\
\>[11]{}\mathbin{<>}{}\<[E]%
\\
\>[11]{}\Varid{tillExit}{}\<[E]%
\\
\>[11]{}(\Varid{branch}\;((\lambda \Varid{xs}\;\Varid{n}\to \Varid{length}\;\Varid{xs}\mathrel{==}\Varid{n})\mathbin{<\hspace{-1.6pt}\mathclap{\raisebox{0.1pt}{\scalebox{.8}{\$}}}\hspace{-1.6pt}>}\mathop{getAll} \mathord{@}\Conid{Int}\;\text{\ttfamily \char34 x\char34}\mathbin{<\hspace{-1.6pt}\mathclap{\raisebox{0.1pt}{\scalebox{.8}{*}}}\hspace{-1.6pt}>}\Varid{getCurrent}\;\text{\ttfamily \char34 n\char34})\;{}\<[E]%
\\
\>[11]{}\hsindent{3}{}\<[14]%
\>[14]{}(\Varid{readInput}\;\text{\ttfamily \char34 x\char34}\;\Varid{ints})\;{}\<[E]%
\\
\>[11]{}\hsindent{3}{}\<[14]%
\>[14]{}\Varid{exit}){}\<[E]%
\\
\>[11]{}\mathbin{<>}{}\<[E]%
\\
\>[11]{}\Varid{writeOutput}\;[\mskip1.5mu \Varid{sum}\mathbin{<\hspace{-1.6pt}\mathclap{\raisebox{0.1pt}{\scalebox{.8}{\$}}}\hspace{-1.6pt}>}\mathop{getAll} \mathord{@}\Conid{Int}\;\text{\ttfamily \char34 x\char34}\mskip1.5mu]{}\<[E]%
\ColumnHook
\end{hscode}\resethooks
We define terms needed for branching and outputs using \ensuremath{\Conid{Applicative}}-style and with the help of the accessors \ensuremath{\Varid{getCurrent}} and \ensuremath{\mathop{getAll} } that correspond to the $C$ and $A$ subscripts of variables.

Now assume we have a program \ensuremath{\Varid{prog}\mathbin{::}\mathit{IO}_\mathit{rep}\;()}.
We can check this program against the specification by running \ensuremath{\Varid{quickCheck}\;(\Varid{prog}\mathbin{`\Varid{fulfills}`}\Varid{spec})}.
This can either result in a successful test run where QuickCheck did not generate a counterexample:
\begin{tabbing}\ttfamily
~~~\char62{}~quickCheck~\char40{}prog~\char96{}fulfills\char96{}~spec\char41{}\\
\ttfamily ~~~\char43{}\char43{}\char43{}~OK\char44{}~passed~100~tests\char46{}
\end{tabbing}
or %
a counterexample is found and we get an error message telling us what went wrong:
\begin{tabbing}\ttfamily
~~~\char62{}~quickCheck~\char40{}prog~\char96{}fulfills\char96{}~spec\char41{}\\
\ttfamily ~~~\char42{}\char42{}\char42{}~Failed\char33{}~Falsifiable\char58{}\\
\ttfamily ~~~Input~sequence\char58{}~\char63{}7~\char63{}2~\char63{}9~\char63{}1~\char63{}\char45{}5~\char63{}1~\char63{}7~\char63{}1\\
\ttfamily ~~~Expected~run~\char40{}generalized\char41{}\char58{}~\char63{}7~\char63{}2~\char63{}9~\char63{}1~\char63{}\char45{}5~\char63{}1~\char63{}7~\char63{}1~\char33{}\char123{}16\char125{}~stop\\
\ttfamily ~~~Actual~run\char58{}~\char63{}7~\char63{}2~\char63{}9~\char63{}1~\char63{}\char45{}5~\char63{}1~\char63{}7~\char33{}15~stop\\
\ttfamily ~~~Error\char58{}\\
\ttfamily ~~~~~AlignmentMismatch\char58{}\\
\ttfamily ~~~~~~~Expected\char58{}~\char63{}1\\
\ttfamily ~~~~~~~Got\char58{}~\char33{}15
\end{tabbing}
Here the program we are testing reads one value less than it should, resulting in a missing input step in its trace.
When checking whether the generalized trace covers the program trace here, we get stuck when checking $\traceOut{15} \traceStop$ against $\traceIn{1} \traceOut{\{16\}} \traceStop$.
This type of mismatch, as already mentioned in Section~\ref{sec:gen-traces}, is one of two possibilities for the coverage check to go wrong in our setting.
The other possible source of a mismatch manifests when the program writes an output that is not part of the set of valid outputs at the respective position.
For such a program, which for example does not include the last read number into the sum, the error message looks like this:
\begin{tabbing}\ttfamily
~~~\char62{}~quickCheck~\char40{}prog~\char96{}fulfills\char96{}~spec\char41{}\\
\ttfamily ~~~\char42{}\char42{}\char42{}~Failed\char33{}~Falsifiable\char58{}\\
\ttfamily ~~~Input~sequence\char58{}~\char63{}3~\char63{}\char45{}2~\char63{}0~\char63{}6\\
\ttfamily ~~~Expected~run~\char40{}generalized\char41{}\char58{}~\char63{}3~\char63{}\char45{}2~\char63{}0~\char63{}6~\char33{}\char123{}4\char125{}~stop\\
\ttfamily ~~~Actual~run\char58{}~\char63{}3~\char63{}\char45{}2~\char63{}0~\char63{}6~\char33{}\char45{}2~stop\\
\ttfamily ~~~Error\char58{}\\
\ttfamily ~~~~~OutputMismatch\char58{}\\
\ttfamily ~~~~~~~the~value~\char45{}2~is~not~covered~by~\char123{}4\char125{}
\end{tabbing}
Note that, as stated in Section~\ref{sec:gen-traces}, there could theoretically also be a mismatch between read values, but since we derive the input sequence from the generalized trace, this cannot occur in our setting.

In both of the error messages shown above, the input sequences all contain only relatively small numbers.
This is not by chance.
When constructing generalized traces, we currently only draw inputs from the range of $-10$ to $10$ for integer and $0$ to $10$ for natural numbers.
This obviously reduces our capabilities to catch certain errors in a program.
On the other hand, since we currently choose elements from these sets completely at random, larger value sets would often result in the generation of very long input sequences or in generation failing to terminate in reasonable time.
This is especially the case if specifications include very particular branching conditions.
Consider, for example, the specification $ (\readInput{x}{\integer} \restCh{\mathit{sum}(x_A) = c} \loopExit)^\loopArr $ for some constant $c$.
In order to reach the desired sum and end the loop, at some point the next input has to be exactly the difference between c and the current sum.
Hitting this one element at random is extremely unlikely when drawing from large sets.
We therefore usually fail to generate even a single trace for such specifications.
Additionally, we are currently not able to give any guarantees when it comes to coverage of branches or other measures of quality for test case distributions.
These are points we are planning to address in the future.

For the same reasons, we do not shrink the found counterexample, like QuickCheck normally does, to obtain a small or even minimal counterexample.
In our example the size of the test case, i.e., of the generalized trace, depends on the first input value, since it determines the number of required executions of the loop body.
Choosing a smaller number for that input would yield a candidate for a smaller counterexample.
But in general, such conditions can be far more complicated.
A possible solution to this problem might be to not rely on random test case generation and shrinking in the first place but instead use an enumerative testing approach~\cite{Duregard2012,Runciman2008}, systematically generating test cases up to a certain size.
However, we currently have not done any concrete work in this direction.

\section{Related work}

The general mechanism for building inspectable representations of side-effecting programs~\cite{swierstra2007} is provided in the Haskell IOSpec library\footnote{\url{https://hackage.haskell.org/package/IOSpec}}.
It supports not only console I/O but also forking processes, mutable references, and software transactional memory.
However, its API is very minimal and no higher-level abstractions currently exist.

Another testing approach that deals with stateful computations is Quviq QuickCheck~\cite{hughes2007,Hughes2016} for Erlang%
\footnote{A Haskell version can be found at \url{http://hackage.haskell.org/package/quickcheck-state-machine}.}.
Instead of testing specific programs, like we do, they test stateful APIs.
A specification of such an API is a semantic model, given in Erlang, of the API together with pre- and post-conditions for each stateful action.
Testing is then done by generating random sequences of actions based on the pre-conditions and checking the result of the actual API calls against the model and post-conditions.
Any found sequence of API calls that do not behave in accordance with the sematic model is simplified via shrinking to obtain a small counterexample.

Due to the large number of existing automatic task grading and assessment tools, we cannot give a complete overview here.
A survey (with a focus on feedback generation) of different automatic assessment tools for programming tasks is presented by Keuning et al.~\cite{keuning2019}.
Most tools use some form of automatic testing, either on specified test cases or by comparing submissions to the results of sample solutions.
Additionally, a number of tools use program transformation or static analysis techniques to determine how a program deviates from a sample solution.
Task specification is usually done through unit or property tests or by providing sample solutions in the respective programming language.
As far as we can tell, no existing system is using a formal specification language for defining intended behavior.
Some systems support automatic generation of exercise tasks, but this is usually restricted to gap-filling tasks derived from sample solutions.

\section{Future work}

\subsection{Improvements to testing}
Our testing framework does not formulate any special search strategy for finding test cases.
It is easy to get stuck while searching for a test case when the random values we choose for inputs do not lead to termination of iterations.
Also, no guarantees on coverage of the complete range of described behavior are given.
Reliable application in a general setting requires that we find solutions for these shortcomings.

\subsection{Moving beyond testing}
Automatic testing of programs is a nice first step when it comes to automating parts of educational activities.
We already have a list of areas in mind for which we would also like some automatization, and our DSL is designed partly with these possibilities in mind:
\begin{itemize}[noitemsep]
  \item
    \textbf{Task generation.}
    We can automatically generate syntactically valid specifications as the basis for new tasks.
    However, we lack a way of controlling the complexity of such specifications.
    The specification language provides ways to control expected input values and some guidance is possible by restricting the term language (via the set $\mathit{Func}$).
    A way of defining and describing meta properties %
    could be used, for example, to provide students with individual tasks while assuring fairness in terms of difficulty etc.
  \item
    \textbf{Sample solutions.}
    It should be fairly obvious that for every expression in our specification language we can automatically generate a (Haskell) program with the respective behavior.
    Unfortunately, such a program is most likely not an ideal sample solution to show to students.
    Manual attempts at transforming such naively generated programs into idiomatic ones suggest that this might be possible in general using basic rewriting and optimization techniques.
  \item
    \textbf{Generation of helpful feedback.}
    Currently the feedback we gain from a failing test case is limited.
    There are no real pointers to the root of the problem but just a basic counterexample for which the program behaves in a certain wrong way.
    Inferring, for example, a specification for the wrongly behaving program and comparing it to the target might reveal the source of an error, enabling us to provide more precise feedback.
  \item
    \textbf{Alternative domains.}
    The general approach of defining a language for specifications from which we can generate black-box tests for free form solutions can potentially be adapted to other programming task domains as well.
    This could include additional I/O capabilities, like reading and writing files.
    But adapting the approach to pure domains like transformations of lists using a certain set of predefined combinators seems possible as well.
    This would allow us to apply automatic task generation etc.\ in pure contexts.
\end{itemize}

\section{Conclusion}

We presented a formal language for specifying the interactive behavior of console I/O programs.
By doing so, we gain the ability to automatically generate tests and test cases to probabilistically check whether a program submission is correct.
This is a significant improvement to our ability to state and automatically grade exercises.
Additionally, the fact that we can manipulate the formal descriptions of behavior programmatically opens up a wide range of possibilities for further automatization and analyses.

\bibliographystyle{eptcs}
\bibliography{sources}

\end{document}